\documentclass[letterpaper, 10 pt, conference]{ieeeconf}  

\IEEEoverridecommandlockouts                              

\usepackage{times}
\usepackage{epsfig}
\usepackage{graphicx}
\usepackage{amsmath}
\usepackage{amssymb}
\usepackage{algorithm}
\usepackage{algorithmic}
\usepackage{mathrsfs}
\usepackage{multirow}

\usepackage{amsthm}
\theoremstyle{plain}

\usepackage[section]{placeins}

\usepackage[]{animate}

\usepackage{graphicx}
\usepackage{animate}

\usepackage{float}

\usepackage{multirow}
\usepackage{rotating}
\usepackage{wrapfig}
\usepackage{lineno}
\usepackage{epsfig}
\usepackage{amsmath,amssymb}
\usepackage{color}
\usepackage{booktabs}       
\usepackage{amsfonts,amsmath}       
\usepackage{nicefrac}       
\usepackage{microtype} 
\usepackage{float}
\usepackage{amsmath}
 
\usepackage{enumerate}

\title{\LARGE \bf
Segmentation of Cardiac Structures via Successive Subspace Learning with Saab Transform from Cine MRI
}

\author{Xiaofeng Liu$^{1}$, Fangxu Xing$^{1}$, Hanna K. Gaggin$^{2}$, Weichung Wang$^{3}$, C.-C. Jay Kuo$^{4}$,~\IEEEmembership{Fellow,~IEEE,} \\Georges El Fakhri$^{1}$,~\IEEEmembership{Fellow,~IEEE,} Jonghye Woo$^{1}$,~\IEEEmembership{Member,~IEEE}
\thanks{}

\thanks{$^{1}$X. Liu, F. Xing, G. El Fakhri, and J. Woo are with the Gordon Center for Medical Imaging, Dept. of Radiology, Massachusetts General Hospital and Harvard Medical School, Boston, MA, USA}

\thanks{$^{2}$H. Gaggin is with Division of Cardiology, Corrigan Minehan Heart Center and Dept. of Medicine, Massachusetts General Hospital and Harvard Medical School, Boston, MA, USA}
\thanks{$^{3}$W. Wang is with Institute of Applied Mathematical Sciences, National Taiwan University, Taipei, Taiwan}

\thanks{$^{4}$C.-C. J. Kuo is with Dept. of Electrical and Computer Engineering, University of Southern California, Los Angeles, CA, USA.}

}

\begin{document}

\maketitle
\thispagestyle{empty}
\pagestyle{empty}

\begin{abstract}

Assessment of cardiovascular disease (CVD) with cine magnetic resonance imaging (MRI) has been used to non-invasively evaluate detailed cardiac structure and function. Accurate segmentation of cardiac structures from cine MRI is a crucial step for early diagnosis and prognosis of CVD, and has been greatly improved with convolutional neural networks (CNN). There, however, are a number of limitations identified in CNN models, such as limited interpretability and high complexity, thus limiting their use in clinical practice. In this work, to address the limitations, we propose a lightweight and interpretable machine learning model, successive subspace learning with the subspace approximation with adjusted bias (Saab) transform, for accurate and efficient segmentation from cine MRI. Specifically, our segmentation framework is comprised of the following steps: (1) sequential expansion of near-to-far neighborhood at different resolutions; (2) channel-wise subspace approximation using the Saab transform for unsupervised dimension reduction; (3) class-wise entropy guided feature selection for supervised dimension reduction; (4) concatenation of features and pixel-wise classification with gradient boost; and (5) conditional random field for post-processing. Experimental results on the ACDC 2017 segmentation database, showed that our framework performed better than state-of-the-art U-Net models with 200$\times$ fewer parameters in delineating the left ventricle, right ventricle, and myocardium, thus showing its potential to be used in clinical practice.   \newline

\indent \textit{Clinical relevance}— Delineation of the left ventricular cavity, myocardium, and right ventricle from cardiac MR images is a common clinical task to establish diagnosis and prognosis of CVD. 
\vspace{+5pt}
\end{abstract}

\section{Introduction}

Cardiovascular disease (CVD) continues to be the cause of the largest portion of morbidity and mortality globally, accounting for over 18 million deaths globally~\cite{roth2020global}. Assessment of CVD with cine magnetic resonance imaging (MRI) has been shown to provide a non-invasive way to evaluate the detailed morphology and function of the heart. In particular, cine MRI is considered to be the most accurate imaging modality for assessing various quantitative parameters with important prognostic implications. 

Segmentation of the left ventricle (LV), right ventricle (RV), and myocardium (MYO) from cardiac cine MR images plays an important role in characterizing clinically important parameters~\cite{gaggin2013biomarkers}, such as ejection fraction (EF), end diastolic volume (EDV), end systolic volume (ESV), and myocardial mass. These parameters, in turn, can be used to identify disease phenotypes, stratify disease risks, and develop diagnostic and prognostic tools \cite{ammar2021automatic}. In clinical practice, semi-automated segmentation is still predominantly used, partly due to the lack of fully-automated and accurate segmentation tools \cite{bernard2018deep}, which is time-consuming and suffers from inter-observer variability. 

With the recent progress of deep learning \cite{goodfellow2016deep}, numerous convolutional neural networks (CNN) models, e.g., U-Net \cite{ronneberger2015u}, have been developed, demonstrating their accuracy in many medical image analysis tasks~\cite{shen2017deep}. While deep learning has achieved impressive results for segmentation and classification, a number of challenges arise in developing and deploying deep learning models for clinical applications \cite{shen2017deep}. First, CNN models typically require a large number of labeled training datasets \cite{goodfellow2016deep}. Sparse and inaccurate labels caused by privacy issues and the high cost of labeling, however, lead to difficulty in collecting sufficient and high-quality training sample datasets \cite{goodfellow2016deep}; with the limited training datasets, an accurate model fitting at the training stage is challenging. Recently, to address this, efforts have been made to generate samples using data augmentation or adversarial training \cite{liu2019hard}, which, however, results in an unavoidable problem of appearance shift between real and generated data. Second, importantly, many CNN models are seen as a ``black-box" model \cite{kuo2016understanding,goodfellow2016deep}. Accordingly, CNN models remain largely elusive how a particular CNN model makes a decision and when it can be trusted. Therefore, it is crucial to develop an explainable model that works with a limited number of datasets for clinical applications.
 
To address the aforementioned challenges, in this work, we propose to develop a lightweight, interpretable, and fully-automated segmentation framework with successive subspace learning (SSL) \cite{rouhsedaghat2021successive}. Specifically, our framework is comprised of the following steps: (1) sequential expansion of near-to-far neighborhood at different resolutions; (2) channel-wise subspace approximation using the subspace approximation with adjusted bias (Saab) transform for unsupervised dimension reduction; (3) a novel class-wise entropy guided feature selection for supervised dimension reduction; (4) concatenation of features and pixel-wise classification with gradient boost; and (5) conditional random field for post-processing.
 
To the best of our knowledge, this is the first attempt at exploring the SSL framework with the Saab transform for a segmentation task. Our framework is lightweight and interpretable, yet achieving a superior segmentation performance with 200$\times$ fewer parameters, compared with state-of-the-art U-Net models.


\section{Methodology}
\subsection{Fundamentals of SSL and Saab Transform}

Inspired by the recent stacked design of CNN models, the SSL principle \cite{rouhsedaghat2021successive} has been targeted for classifying 2D natural images (e.g., PixelHop \cite{chen2020pixelhop,zhang2020pointhop}), 3D MR images \cite{liu2021voxelhop}, and point clouds (e.g., PointHop \cite{zhang2020pointhop}). In each layer of SSL, the Saab transform \cite{kuo2019interpretable}, a variant of Principal Component Analysis (PCA), is used as an alternative to nonlinear activation, thereby alleviating the sign confusion problem \cite{kuo2016understanding}. Furthermore, the Saab transform is deemed more interpretable than nonlinear activation functions in CNNs \cite{kuo2019interpretable,fan2020interpretability}, as the model parameters are computed stage-by-stage in a feedforward manner, without backpropagation. Accordingly, the training of our SSL-based method is more efficient and interpretable than that of CNN models \cite{chen2020pixelhop}.

 

\subsection{Our Saab-based SSL segmentation Framework}

\begin{figure}[t]
  \centering\vspace{+5pt}
\includegraphics[width=8.8cm]{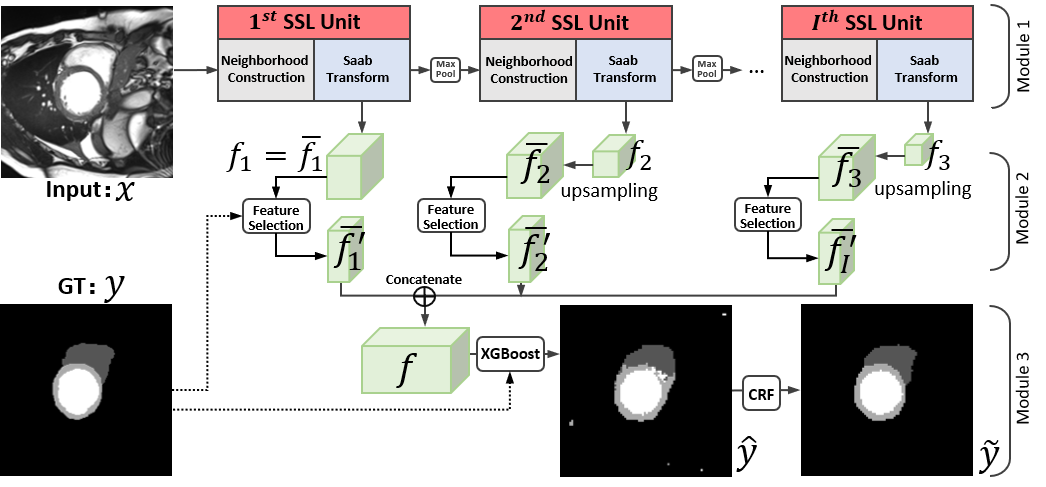}  \vspace{-15pt}
  \caption{Illustration of our proposed framework using  the Saab transform, which consists of 3 modules.}\label{fig1} 
\end{figure}

In this work, we have a 2D MR image ${\bf x}\in\mathbb{R}^{H\times W\times 1}$ and its corresponding label ${\bf y}\in\mathbb{R}^{H\times W\times 4}$, where $H$ and $W$ denote the horizontal and vertical dimensions, respectively. The channel of the gray-value sample is 1, and the label of each pixel is encoded as a four-dimensional one-hot vector for four tissue classes. The architecture of our framework is illustrated in Fig.~\ref{fig1}, as detailed below.

\subsubsection{Module 1: Unsupervised Feature Selection}

We first construct $I$ cascade SSL units and $I-1$ max-pooling operations to extract the attributes at different spatial scales in the unsupervised Module 1.  
Similar to PixelHop \cite{chen2020pixelhop}, in each SSL unit, we construct the neighboring region on the $H\times W$ plane. For instance, in the first SSL unit, for the single-channel data, we construct the $3\times3$ region for each pixel position. Each of them is then flattened to a 9-dimensional vector. With a padding operation, {\bf x} is transformed to a cubic with the size of $H\times W\times 9$. Then, the Saab transform is used for unsupervised dimension reduction in the channel direction. Each 9-dimensional vector is mapped to a $F_1$-dimensional feature vector, where $F_1$ is a hyperparameter to control the output dimension of the first PixelHop unit. 

Specifically, the terms, direct current (DC) and alternating current (AC), are adopted from the circuit theory. In the first Saab transform, we configure one DC and $F_1-1$ AC anchor vectors with the size of $H\times W \times 1$. Then, the $c$-th dimension of $f$ can be an affine transform of $x$, i.e., 
\begin{equation}
    f_c=a_c^Tx+b_c,~~~ c=0,1,\cdots, F_1-1,
\end{equation} 
and the Saab transform has a special design of the anchor vector $a_c\in\mathbb{R}^{1\times(H\times W \times 1)}$ and the bias term $b_c\in\mathbb{R}$ \cite{kuo2019interpretable}. Similar to \cite{kuo2019interpretable}, we can set $b_c\equiv d\sqrt{F_1}, d\in\mathbb{R}$, and divide the anchor vector into two categories:
\begin{equation}
\begin{aligned}
&\bullet~{\text{DC anchor vector}}~~a_0=\frac{1}{\sqrt{H\times W \times 1}}(1,\cdots, 1)^T, \\
&\bullet~{\text{AC anchor vector}}~~a_c,~~c=1,\cdots, F_1-1. \label{eq:1}
\end{aligned}\end{equation}

After computing $f_1\in\mathbb{R}^{H\times W\times F_1}$, we half its spacial size with the max-pooling operation to ${\frac{H}{2}\times \frac{W}{2}\times F_1}$ and send to the next SSL unit. With the multi-channel input, the neighborhood construction involves $3\times 3\times F_1$ region at each pixel position. Then, the neighborhood union is flattened to a vector, which is further processed by the Saab transform for dimension reduction. The detailed structure of our module 1 is provided in Table \ref{table:1}.   

With the cascaded SSL units, the neighborhood union is correlated with more pixels of $\bf x$ to extract global information. This process is similar to CNN models in that a larger reception field is achieved in the deeper layers. 

\subsubsection{Module 2: Supervised Feature Selection}
In what follows, we resort to the supervised dimension reduction based on class-wise entropy-guided feature selection to tailor the discriminative feature for our segmentation task. 

Because of the resolution deduction in each unit, we have different spatial size of the extracted features. The features in the later units correspond to a larger reception field (i.e., more pixels) in ${\bf x}$ and ${\bf y}$. To match the features in these units with the original pixels, we resize $f_2,\cdots,f_I$ to the size of $f_1$ and denote as $\overline{f_2},\cdots,\overline{f_I}$. Therefore, we have $\overline{f_1}\in\mathbb{R}^{H\times W\times C_1}$, $\overline{f_2}\in\mathbb{R}^{H\times W\times C_2}$, $\cdots$, $\overline{f_I}\in\mathbb{R}^{H\times W\times C_I}$. 

Because of the disparate importance, depending on the different channels for the segmentation decision, it is necessary to make supervised feature selection. In related developments, PixelHop++ \cite{chen2020pixelhop++} proposes to classify each channel with the size of $\mathbb{R}^{H\times W}$ and select the channels with low cross-entropy score. However, it is not applicable to segmentation as a channel selection, since the label in the segmentation task is pixel-wise and the feature of a pixel in each channel is only a scalar, making it challenging to be used as a feature for a classifier.

Instead, we propose to select the channel with the small entropy of each class. Specifically, we would encourage the feature of a pixel in each channel to be similar, if the label of the corresponding pixels is the same class. We denote the feature of a pixel in each channel $p_i^c$ for the $i$-th pixel of a class in the $c$-th channel. The entropy of a sample can be:
\begin{equation}
\begin{aligned}
H=\sum_{j=1}^4 H_j,~~H_j= -\sum_{i} p_i^c {\text{log}} p_i^c, \label{eq:1}
\end{aligned}\end{equation}
where we use $j$ to index the four classes in our segmentation task. After calculating the entropy of four classes for each channel, we rank the entropy in descending order. Then, we select the top 80\% channels for the subsequent pixel-wise classification task.

\begin{table}[t]
\caption{The detailed structure of our 4 consecutive SSL units} 
\centering 
\resizebox{1\columnwidth}{!}{%
\begin{tabular}{l | l | l} 
\hline\hline 
Input Size&Type& Filter Shape   \\ [0.5ex] 
\hline 

$[224\times224\times1]$&Saab Trans& $F_1$ kernels of $3\times3$\\
$[224\times224\times F_1]$&MaxPool& (2$\times$2)-(1$\times$1)\\
\hline

$[112\times112\times F_1]$&Saab Trans& $F_2$ kernels of $3\times3$ for F1 channels\\
$[112\times112\times F_2]$&MaxPool& (2$\times$2)-(1$\times$1)\\
\hline

$[56\times56\times F_2]$&Saab Trans& $F_3$ kernels of $3\times3$ for F2 channels\\
$[56\times56\times F_3]$&MaxPool& (2$\times$2)-(1$\times$1)\\
\hline

$[28\times28\times F_3]$&Saab& $F_4$ kernels of $3\times3$ for F3 channels\\
\hline

\end{tabular}
}\label{table:1}
\end{table}

\subsubsection{Module 3: Information fusion for segmentation and post-processing}

With the extracted features $\overline{f_2'},\cdots,\overline{f_I'}$ with both the Saab transform and class-wise entropy guided selection, we concatenate them along with the channel dimension to get the feature $f\in\mathbb{R}^{H\times W\times C}$. The channel dimension $C$ is the sum of all channels in $\overline{f_2'},\cdots,\overline{f_I'}$. Each feature vector on the $H\times W$ plane of $f$ corresponding to an original pixel in ${\bf x}$ or ${\bf y}$. Then, we carry out the pixel-wise classification for each of $C$ dimensional features with a classifier. We empirically choose the extreme gradient boosting (XGBoost) \cite{chen2015xgboost}, which is an optimized distributed gradient boosting library designed to be highly efficient and flexible. XGBoost is trained to learn the correlation of pixel-wise feature and ground truth pixel class label in our training set. 

We note that with a limited number of SSL units, it is challenging to support the reception field to cover all of the pixels for global perception. In contrast, too many SSL units will lead to very low resolution in the later units, which is not sufficient to support the pixel-wise segmentation. In addition, the channel size of the later units will be very large, leading to a long and indiscriminative feature vector, which can distract the pixel-wise classification. 

To balance this conflict, we propose to validate the most effective $I$ and adopt the well-established post-processing tool of conditional random field (CRF) to further refine the segmentation results and get the final results of our framework $\tilde{y}=CRF(\hat{y})$.

\section{Experiments}


To demonstrate the performance of our Saab transform-based SSL framework, we validated it on the Automated Cardiac Diagnosis Challenge (ACDC 2017) database, which contains 100 subjects. The cine MRI short-axis slices were acquired with 1.5T or 3.0T MRI scanners. The acquired cine MRI short-axis slices covered the LV, RV, and MYO from the base (upper slice) to the apex (lower slice), with 5–8 mm slice thickness, 5 or 10 mm inter-slice gap and the spatial resolution of 1.37–1.68 $mm^2$.

\begin{figure}[t!]
  \centering 
\includegraphics[width=8.8cm]{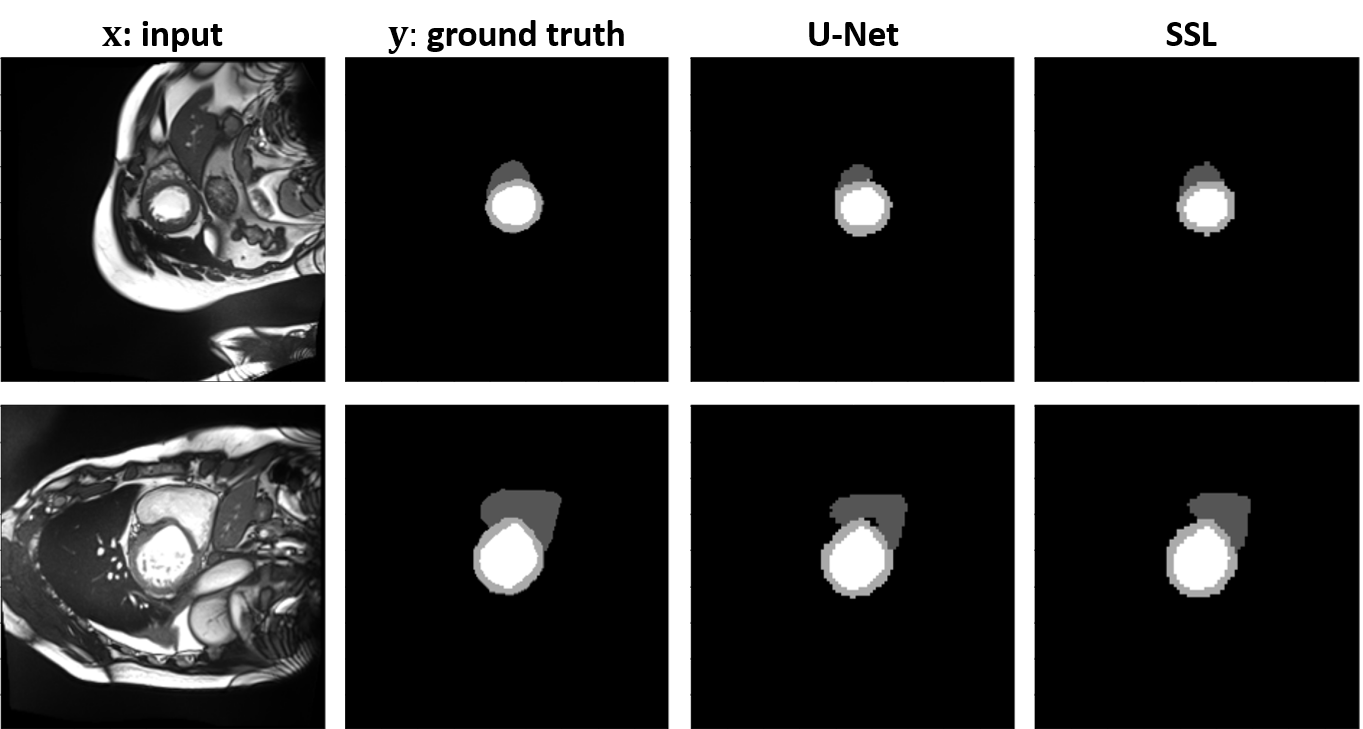} \vspace{-10pt}
  \caption{Comparison of the segmentation results with 60 subjects for training.}\label{figexp1}  \vspace{+5pt}
\end{figure}

 \begin{table}[t!]
\caption{Comparison of the Dice score with 50 training subjects}\vspace{-5pt} 
\centering 
\resizebox{1\columnwidth}{!}{%
\begin{tabular}{l | c | c   c   c | c} 
\hline\hline 
Methods& \textbf{Parameters}  & RV & MYO & LV & \textbf{Average} \\  
\hline 

U-Net&5.88 M & 81.45\% & 78.54\% & 90.07\% & 83.35\% \\
AttenUNet&6.40 M & 81.02\% & 78.40\% & 89.32\%&82.91\% \\\hline
SSL&\textbf{0.03 M}& \textbf{82.91\%} & \textbf{81.83\%} & \textbf{90.62\%}& \textbf{85.12\%}  \\\hline
\end{tabular}
}\label{table:2}
\end{table}

\begin{table}[t!]
\caption{Comparison of the Dice score with 60 training subjects} \vspace{-5pt}
\centering 
\resizebox{1\columnwidth}{!}{%
\begin{tabular}{l | c | c   c   c | c} 
\hline\hline 
Methods& \textbf{Parameters}  & RV & MYO & LV & \textbf{Average}\\  
\hline 

U-Net&5.88 M & 85.54\% & 78.81\% & \textbf{92.15\%} & 85.50\% \\
AttenUNet&6.40 M & \textbf{85.64\%} & 77.37\% & 91.29\% & 84.76\%\\\hline
SSL&\textbf{0.03 M}& 83.89\% & \textbf{82.57\%} &  {91.62\%} & \textbf{86.03\%} \\\hline
\end{tabular}
}\label{table:3}
\end{table}

For each patient, the delineations of the LV, RV, and MYO, were obtained by two clinical experts. On average, each subject had about 27 labeled slices. We reported the average Dice similarity score with 30 subjects for testing, 10 subjects for validation, and 50 or 60 subjects for training.

\subsection{Implementation Details}
All the experiments were implemented using Python on a server with a Xeon E5 v4 CPU/Nvidia Tesla V100 GPU with 128GB memory. We also used the widely adopted deep learning library, Pytorch, to implement U-Net \cite{ronneberger2015u} and AttnUNet \cite{schlemper2019attention}. For a fair comparison, we resized all of the slices to $224\times 224\times 1$, which was consistent with the input of the U-Net models.

We empirically used four SSL units and set $F_1$=5, $F_2$=10, $F_3$=30, and $F_4$=100. We note that the number of the Saab AC filters in the unsupervised dimension reduction procedure controls the preserved energy ratio.

\subsection{Experimental Results}

Fig. \ref{figexp1} shows the segmentation results of U-Net with ResNet50 backbone and our SSL framework. We can see that SSL is able to achieve comparable or even better performance than the widely used U-Net models. 

For quantitative evaluation, we compared the Dice similarity score in Tables \ref{table:2} and \ref{table:3}, which used 50 or 60 subjects for training, respectively. Note that the larger Dice similarity score indicates the better segmentation performance. The best results are bolded. With 50 subjects for training, our SSL framework outperformed U-Net \cite{ronneberger2015u} and attention-based U-Net \cite{schlemper2019attention} in all of the three classes. We can observe that with relatively limited training datasets, the performance of the CNN models is inferior to our framework. In addition, the statistics of the network parameters are provided and compared in Table \ref{table:2}. We can see that the number of parameters of our SSL framework was about 200 times fewer than the popular U-Net structures \cite{ronneberger2015u,schlemper2019attention}. The much fewer parameters can largely alleviate the difficulty of a small number of training datasets. In the case of using 60 subjects for training, our SSL framework achieved better performance than the U-Net based methods in the average Dice similarity score.

\subsection{Sensitivity Analysis and Ablation Study}

\begin{figure}[t]
  \centering\vspace{+5pt}
\includegraphics[width=8cm]{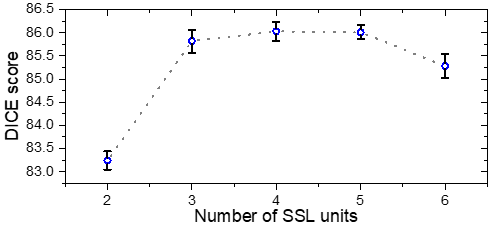}  \vspace{-5pt}
  \caption{Sensitivity analysis of the number of SSL units for the case of using 60 subjects in training.}\label{figexp3} 
\end{figure} 

\begin{figure}[t]
  \centering\vspace{+5pt}
\includegraphics[width=8.8cm]{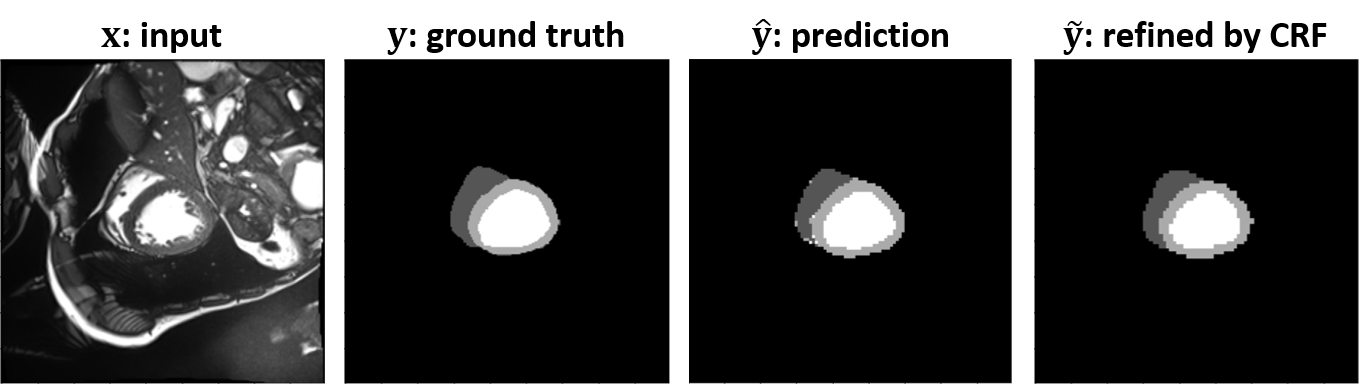}  \vspace{-10pt}
  \caption{Comparison of with or without CRF post-processing.}\label{figexp222} 
\end{figure}

With four SSL units, we achieved a state-of-the-art Dice similarity score in both 50 and 60 training subjects settings. The number of SSL units is important for our segmentation framework to balance the efficiency and perception area. The low resolution can be challenging to provide accurate information for fine-grained pixel-wise classification. In Fig. \ref{figexp3}, we have shown the detailed sensitivity study using different SSL units. The standard deviation was computed with five random choices of training and validation splits. The class-wise entropy-guided feature selection was developed to simplify the subsequent classification modules. In addition, CRF was applied as a post-processing step. To demonstrate their effectiveness, we provide the ablation study in Table \ref{table:4} and the effect of CRF in Fig. \ref{figexp222}.



\begin{table}[t]
\caption{Results of Ablation Studies with 60 training subjects}\vspace{-5pt} 
\centering 
\resizebox{0.9\columnwidth}{!}{%
\begin{tabular}{l | c} 
\hline\hline 
Methods&Average DICE  \\  
\hline 
SSL&\textbf{86.03\%}\\\hline
SSL without CRF& 84.76\%\\
SSL without entropy-guided feature selection& 85.91\%\\
\hline
\end{tabular}
}\label{table:4}
\end{table}

\section{Conclusion}

In this work, we presented a lightweight, interpretable, and fully-automated SSL framework with the Saab transform to segment the LV, RV, and MYO from cine MRI. A novel class-wise entropy-guided feature selection was proposed to achieve accurate segmentation. Our thorough experiments carried out using the ACDC 2017 database with different number of training subjects demonstrated that our framework achieved a superior performance, compared with the U-Net-based approaches, with about 200$\times$ fewer parameters.

\bibliographystyle{IEEEtran}
\bibliography{main}

\end{document}